\documentclass[pdflatex,sn-mathphys-num,oneside]{sn-jnl}
\makeatletter
\@twosidefalse
\@mparswitchfalse
\makeatother

\usepackage{graphicx}%
\usepackage{multirow}%
\usepackage{amsmath,amssymb,amsfonts}%
\usepackage{amsthm}%
\usepackage{mathrsfs}%
\usepackage[title]{appendix}%
\usepackage{xcolor}%
\usepackage{textcomp}%
\usepackage{manyfoot}%
\usepackage{booktabs}%
\usepackage{tikz}
\usepackage{algorithm}%
\usepackage{algorithmicx}%
\usepackage{algpseudocode}%
\usepackage{listings}%
\usepackage{placeins}%

\theoremstyle{thmstyleone}%
\theoremstyle{thmstyletwo}%

\theoremstyle{thmstylethree}%

\begin{document}

\title{Quantum Hash Function Based on Spectral Properties of Graphs and Discrete Walker Dynamics}

\author*[1]{\fnm{Mohana Priya} \sur{Thinesh Kumar}}\email{manthramohana1@gmail.com}
\equalcont{These authors contributed equally to this work.}
\author*[1]{\fnm{Pranavishvar} \sur{Hariprakash}}\email{pranavishvar@gmail.com}
\equalcont{These authors contributed equally to this work.}



\affil*[1]{\orgdiv{Department of Physics}, \orgname{Indian Institute of Technology (Indian School of Mines), Dhanbad}, \orgaddress{\postcode{826004}, \state{Jharkhand}, \country{India}}}




\abstract{We present \textit{Quantum Graph Hash} (QGH-256), a novel quantum spectral hashing
algorithm that generates high-entropy fingerprints from message-induced graphs.
Each input message is mapped to a weighted graph via a discrete random walk on
an $n\times n$ toroidal grid, where the walk dynamics determine the edge
weights. Quantum Phase Estimation (QPE) is then used to extract the phase
spectrum of the graph Laplacian. Unlike standard QPE settings, the phase
estimation is performed with respect to a superposition state (a uniform
superposition over all node basis states) rather than an eigenvector, ensuring
that all eigencomponents contribute to the resulting spectrum. This yields
spectral features that distinguish even co-spectral but non-isomorphic
message-induced graphs. The final spectral fingerprint is converted into a 256-bit digest, producing a
compact representation of the input. As the fingerprint encodes both spectral
and dynamical properties of the message-induced graph, the resulting hash
exhibits strong sensitivity to input perturbations and provides a structurally
rich foundation for post-quantum hashing. To demonstrate the feasibility of the approach, we implement QGH-256 
on a $4\times 4$ toroidal grid, chosen empirically: smaller grids exhibit 
collisions, whereas larger grids significantly increase execution time. The 
entire pipeline is implemented in Qiskit, and we use a seeded statevector 
simulator to obtain stable, noise-free results.
}




\maketitle

\section{Introduction}\label{sec1}

As the digital landscape continues its rapid evolution, ensuring the security and integrity of sensitive data remains a paramount challenge. Cryptographic hash functions, fundamental constructs in modern cryptography since their early formulations, play an indispensable role in data integrity verification by transforming messages of arbitrary length into fixed-length digests While classical standards like MD5, SHA-1, and SHA-256 have historically served as foundational primitives, they increasingly face vulnerabilities, including collision susceptibility and weakened resistance against advanced computational attacks. 
\vspace{-0.5cm}
\paragraph{}
The advent of quantum computing introduces an existential threat to these classical cryptographic mechanisms. Algorithms such as Shor's, capable of efficiently factoring large integers, and Grover's, which can accelerate unstructured searches, exploit quantum parallelism to compromise the security assumptions underlying many existing hash functions. These quantum threats necessitate the development of post-quantum and quantum-native cryptographic designs that can withstand both classical and quantum attacks, a need underscored by initiatives like the U.S. National Institute of Standards and Technology Post-Quantum Cryptography (PQC) Standardization process.
In response to these critical challenges, we present a novel quantum spectral hashing algorithm designed to generate high-entropy and robust cryptographic fingerprints. Our approach integrates concepts from classical walk theory and quantum spectral graph analysis within a hybrid classical-quantum computational model. 
\vspace{-0.5cm}
\paragraph{}
The paper is organized as follows. Section~\ref{sec2} reviews the prerequisite background on classical discrete walks in one and two dimensions. Section~\ref{sec3} introduces the message-induced weighted graph construction that underlies our hashing framework. Section~\ref{sec4} presents the spectral fingerprinting mechanism. Section~\ref{sec5} integrates these components into the complete QGH-256 workflow. Section~\ref{sec6} discusses evaluation of the cryptographic properties of QGH-256. Section~\ref{sec7} examines potential applications of QGH-256. Section~\ref{sec8} discusses quantum threats to classical cryptography and analyzes the compliance of QGH-256 with NIST PQC standards. The paper concludes by highlighting the core contributions of QGH-256, its evaluation using the seeded simulator, and the planned future work.

\section{Prerequisites}\label{sec2}

\subsection{One-Dimensional Classical Random Walk}
\paragraph{} 
Consider a person standing on a straight line, able to move only in one dimension. 
The direction of the person's movement is determined by repeatedly tossing a coin. 
If the outcome of the toss is a head, the person moves one step forward; if the 
outcome is a tail, the person moves one step backward. After each step, the person 
tosses the coin again to determine the next move. This simple mechanism forms the 
core idea of a classical random walk.

\begin{figure}[h!]
    \centering
    \includegraphics[width=1.0\textwidth]{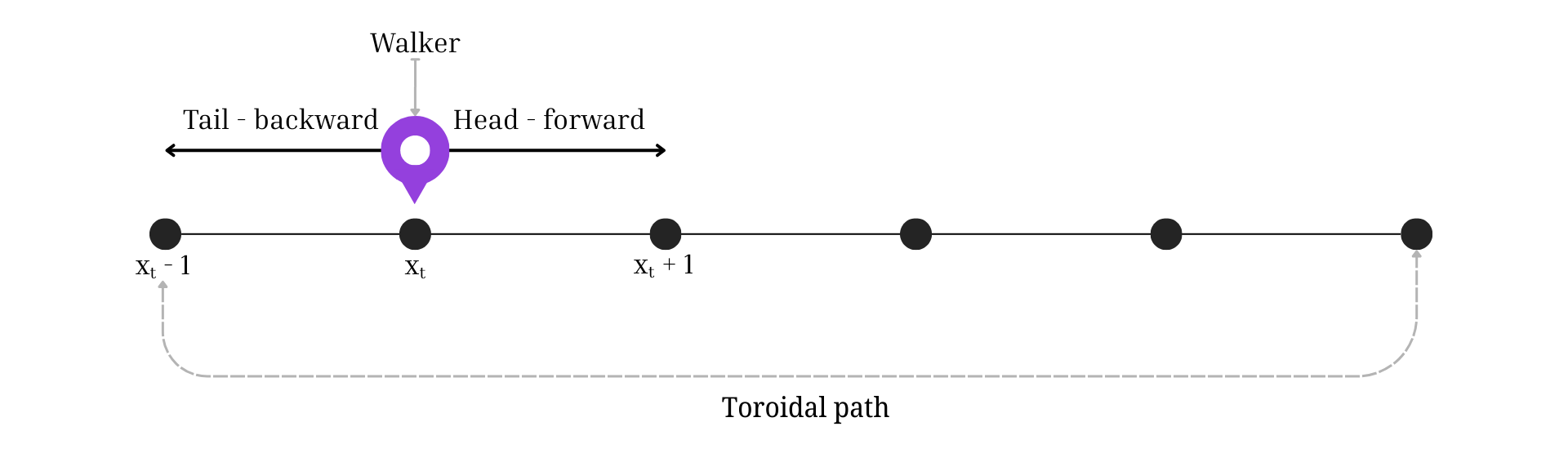}
    \caption{One-dimensional classical random walk along the toroidal path}
    \label{fig:spectral_fingerprint}
\end{figure}
\vspace{-1cm}

\paragraph{}
The position update rule for each time step can be written as
\[
x_{t+1} = 
\begin{cases}
x_t + 1, & \text{if the coin shows head (H)},\\[6pt]
x_t - 1, & \text{if the coin shows tail (T)}.
\end{cases}
\]

\paragraph{} 
Thus, the motion of the walker is entirely governed by the outcomes of independent 
coin tosses. Over many steps, this process generates a trajectory consisting of 
forward and backward movements, producing a random spreading pattern around 
the starting point.

\subsection{Two-Dimensional Classical Random Walk}

\paragraph{}
We now extend the above process to higher dimensions. Consider a person standing on 
a two-dimensional plane. Unlike the one-dimensional walker that uses a single coin, 
the two-dimensional walker uses two coins to determine each movement step. The coins 
are not tossed simultaneously; instead, they are tossed sequentially, and hence the 
order of outcomes plays a crucial role in determining the direction of motion. 
\vspace{-0.5cm}
\paragraph{}
Let the outcome of the first coin be the first bit and the outcome of the second 
coin be the second bit in a two-bit sequence. Each ordered pair of outcomes 
corresponds uniquely to one of the four possible directions:
\[
\begin{aligned}
TT &\;\rightarrow\; \text{up (North)},\\
HH &\;\rightarrow\; \text{forward (East)},\\
HT &\;\rightarrow\; \text{down (South)},\\
TH &\;\rightarrow\; \text{backward (West)}.
\end{aligned}
\]

\vspace{-0.5cm}
\begin{figure}[h!]
    \centering
    \includegraphics[width=1\textwidth]{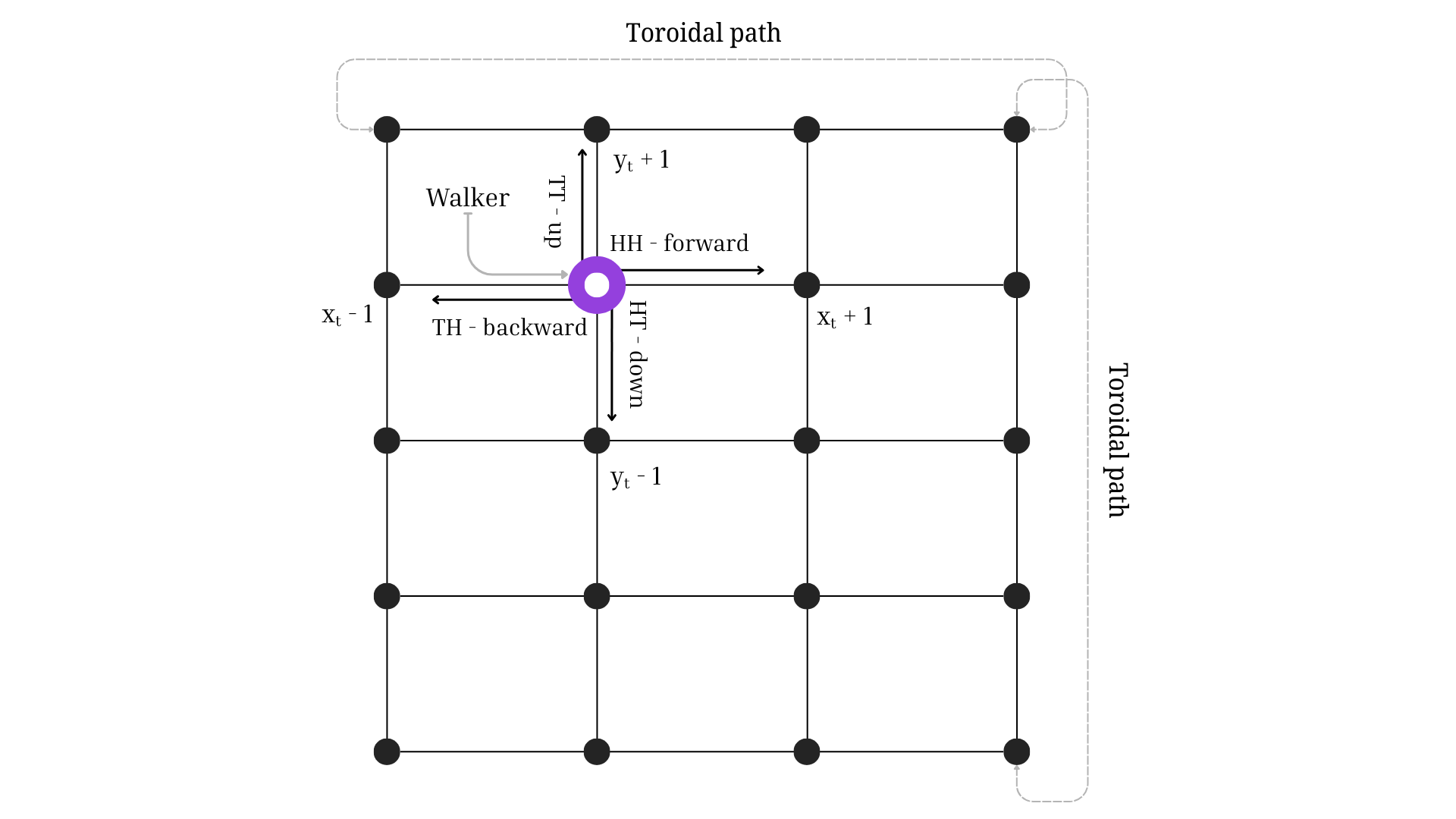}
    \caption{Two-dimensional classical random walk along the toroidal path}
    \label{fig:spectral_fingerprint}
\end{figure}
\vspace{-0.5cm}

\paragraph{}
Since both coins are fair and independent, the walker has four equally likely 
directions to move at each step. Therefore, the movement probabilities are
\[
P(\text{North}) = P(\text{South}) = P(\text{East}) = P(\text{West}) = \frac{1}{4}.
\]

\paragraph{}
Based on the two-bit outcome obtained from the sequential coin tosses, the position 
update rule is
\[
(x_{t+1},\, y_{t+1}) =
\begin{cases}
(x_t,\, y_t + 1), & TT \;\text{(move North)}, \\[6pt]
(x_t,\, y_t - 1), & HT \;\text{(move South)}, \\[6pt]
(x_t + 1,\, y_t), & HH \;\text{(move East)},  \\[6pt]
(x_t - 1,\, y_t), & TH \;\text{(move West)}.
\end{cases}
\]

\paragraph{}
As in the one-dimensional case, the walker repeats this procedure at every step. 
Over time, the sequence of ordered coin outcomes generates a path that spreads 
throughout the plane, producing the classical two-dimensional random walk. The 
sequential nature of the coin tosses ensures that the order of outcomes uniquely 
determines the direction, preventing any ambiguity in the movement rule.

\section{A Unique Graph For Every Message Using
Discrete Walker}\label{sec3}
\paragraph{Message-Induced Walker Model.}

\paragraph{}
In this work, we propose a procedure that assigns a unique weighted map to every input 
message. The walker begins on a fixed \(4\times 4\) grid, and unlike the classical 
random walk where movement is chosen randomly, here the walker's direction at each 
step is entirely determined by the message itself.
\vspace{-0.5cm}
\paragraph{}
We first convert the input message into its UTF--8 binary representation. The 
resulting bitstring is then divided into consecutive blocks of two bits. If the 
total number of bits is odd, a single zero is appended to the end of the string to 
ensure that an even number of bits is available for pairing. Thus the message is 
represented as
\[
M = \{\,b_1b_2,\; b_3b_4,\; \dots,\; b_{n-1}b_n\,\},
\]
\vspace{-1.0cm}
\paragraph{}
where each two-bit block prescribes one step of the walk. Each block is mapped to one of the four possible movement directions according to a 
fixed one-to-one correspondence:
\[
00 \rightarrow \text{down}, \qquad
01 \rightarrow \text{up},   \qquad
10 \rightarrow \text{right},\qquad
11 \rightarrow \text{left}.
\]
This mapping may be modified as needed to suit the preferences of the encoder or could be made more secure by making the mappings keyed. 
\vspace{-0.5cm}
\paragraph{}
Starting from the initial grid position, the walker reads the two-bit blocks in 
order and moves accordingly. Every time the walker traverses an edge, the 
corresponding weight on that edge is incremented by one. As the walker continues to 
trace the path dictated by the message, a weighted graph is formed. The final output 
is a weighted graph
\[
G = (V, E, W),
\]
\vspace{-1cm}
\paragraph{}
where \(V\) is the set of grid nodes, \(E\) is the set of edges of the grid, and 
\(W : E \to \mathbb{N}\) assigns to each edge its cumulative traversal count.

\begin{figure}[h!]
    \centering
    \includegraphics[width=0.9\textwidth]{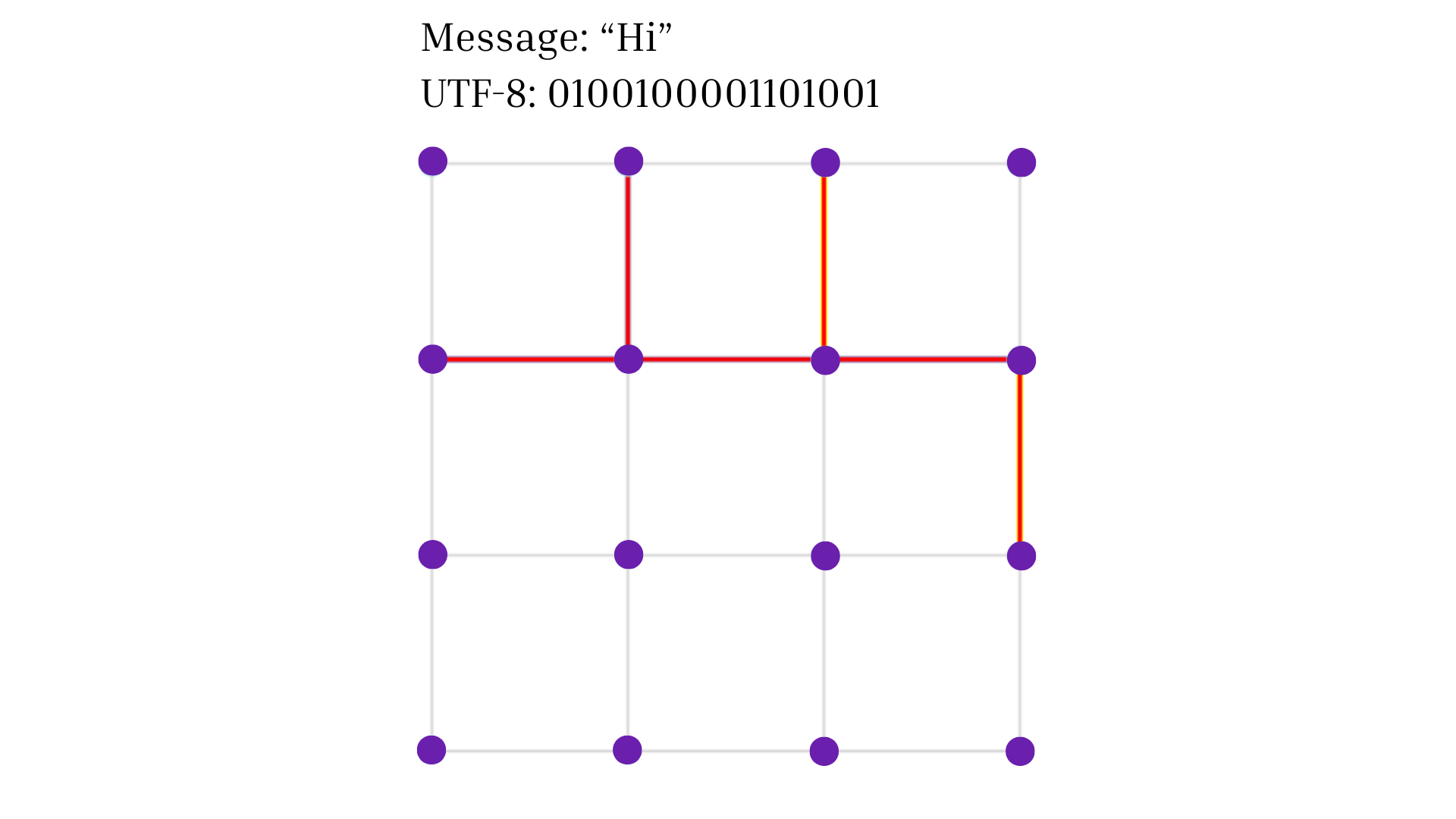}
    \caption{Message induced discrete walker path traced in 4X4 toroidal grid for the message "Hi"}
    \label{fig:spectral_fingerprint}
\end{figure}

\FloatBarrier
\vspace{-0.5cm}
\paragraph{}
Each input message produces a distinct and reproducible weighted map: the same 
message will always generate the same weighted graph. 

\vspace{-0.5cm}
\paragraph{}
\label{msg_from_graph}
However, recovering the 
original message from the resulting graph is computationally hard. Because the graph 
is undirected, the edge weights record only the total number of times each edge was 
traversed, without any information about the direction of traversal. Additionally, 
without knowledge of the walker's final node or the sequence in which edges were 
visited, reconstructing the exact path becomes extremely difficult. This problem 
becomes even harder if the mapping between two-bit blocks and movement directions is 
keyed or randomized in a message-dependent manner through additional processing. 
Under such conditions, even with full access to the weighted graph, determining the 
original message is effectively infeasible.

\section{Spectral Fingerprint of a Graph}\label{sec4}

The spectral fingerprint of a graph is a process through which we generate a unique 
bitstring associated with a given graph using the quantum algorithm known as Quantum 
Phase Estimation (QPE). Just as every human fingerprint is distinct, the spectral 
fingerprint of a graph captures its intrinsic structural properties in a way that 
no two non-isomorphic graphs share the same spectrum under ideal conditions. In 
this section, we outline the steps required to obtain this spectral representation.

\subsection{Hamiltonian Encoding of the Graph Laplacian}

We begin by encoding the input graph using its Laplacian matrix \(L\), which 
incorporates both degree information and connectivity between vertices. The 
standard Laplacian of a graph is defined as
\[
L = D - A,
\]
where \(D\) is the degree matrix and \(A\) is the adjacency matrix. A graph's degree matrix is a diagonal matrix where each diagonal entry is the degree of the corresponding vertex, and all off-diagonal entries are zero. The degree of a vertex is the number of edges connected to it.
An adjacency matrix of a weighted graph is a square matrix where each cell \((i,j)\) contains the weight of the edge between vertex \(i\) and vertex \(j\).

\paragraph{Example:}  
\begin{figure}[h!]
    \centering
    \includegraphics[width=1.0\textwidth]{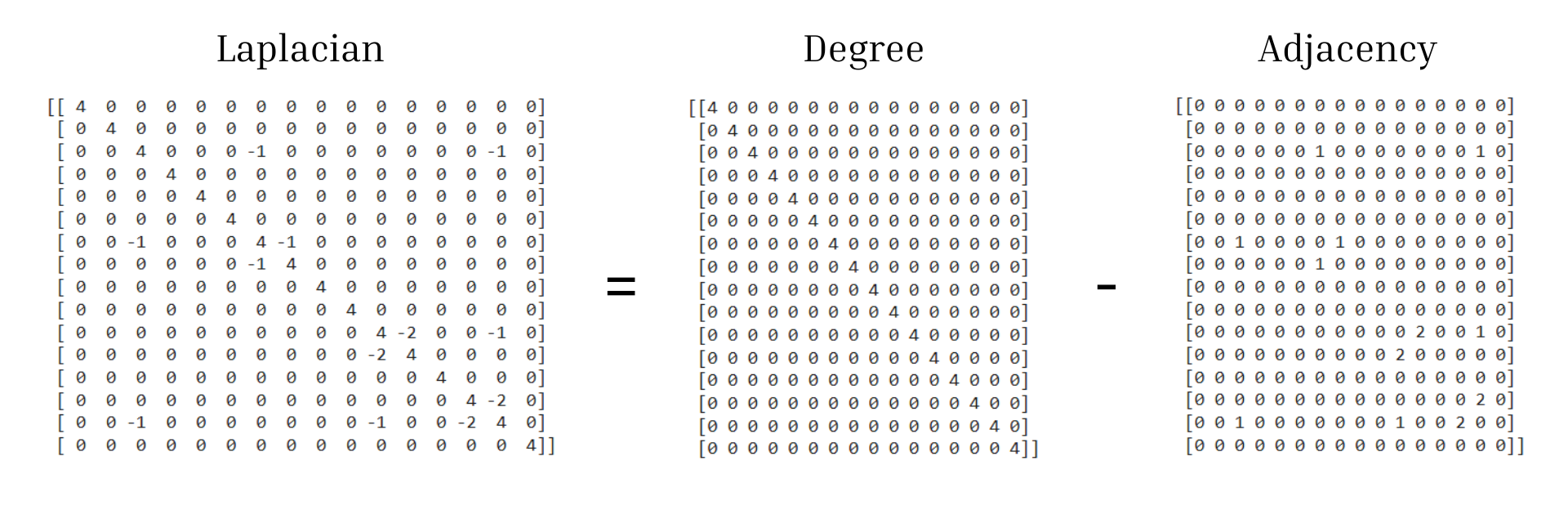}
    \caption{Graph Laplacian of the weighted graph produced by the message "Hi"}
    \label{fig:spectral_fingerprint}
\end{figure}

\vspace{-0.5cm}

\paragraph{Hermiticity.}
A Hermitian matrix is a square complex matrix that is equal to its own conjugate transpose.For a real matrix, being Hermitian is equivalent to being symmetric (\(A^{T}=A\)).
The Laplacian \(L\) of any undirected graph is real and symmetric, and therefore Hermitian:
\[
L^\dagger = L.
\]
\vspace{-0.5cm}
\paragraph{}
However, \(L\) itself is \emph{not} unitary. To use it in quantum phase estimation, 
we must convert it. This is where Matrix exponentiation comes into play.

\subsection{Unitary Transformation via Matrix Exponentiation}

To convert the Hermitian Laplacian matrix $L$ (where $L = L^\dagger$) into a Unitary matrix $U$ suitable for quantum computation, we utilize matrix exponentiation. The transformation is defined as:

\[
U = e^{iLt}
\]

\vspace{-1.0cm}
\paragraph{}
Here, $t$ acts as a time-evolution parameter. 

\subsubsection{Proof of Unitarity}
First, we take the conjugate transpose of the exponential definition:
\[
U^\dagger = (e^{iLt})^\dagger = e^{-iL^\dagger t}
\]

\vspace{-1.0cm}
\paragraph{}
Since $L$ is Hermitian ($L^\dagger = L$), this simplifies to:
\[
U^\dagger = e^{-iLt}
\]

\vspace{-1.0cm}
\paragraph{}
Multiplying $U^\dagger$ by $U$:
\[
    U^\dagger U = e^{-iLt} \cdot e^{iLt} = e^{(-iLt + iLt)} = e^0 = I
\]

\vspace{-0.5cm}
\paragraph{}
Since the product is the Identity matrix $I$, the operator $U$ is unitary.

\subsubsection{Eigenvalue Transformation}

A critical property of matrix exponentiation is how it acts on the eigenbasis. While the operation \textbf{preserves the eigenvectors}, it \textbf{exponentiates the eigenvalues}.
\vspace{-0.5cm}
\paragraph{}
If $|v\rangle$ is an eigenvector of $L$ with a real eigenvalue $\lambda$ (i.e., $L|v\rangle = \lambda|v\rangle$), we can determine the effect of $U$ using the Taylor series expansion:

\begin{align*}
    U |v\rangle &= e^{iLt} |v\rangle \\
    &= \left( \sum_{k=0}^{\infty} \frac{(iLt)^k}{k!} \right) |v\rangle \\
    &= \left( \sum_{k=0}^{\infty} \frac{(i \lambda t)^k}{k!} \right) |v\rangle \\
    &= e^{i \lambda t} |v\rangle
\end{align*}
\vspace{-1.0cm}
\paragraph{}
This step works because $|v\rangle$ is an eigenvector of the operator $L$, 
with eigenvalue $\lambda$. That is:

\[
L|v\rangle = \lambda |v\rangle.
\]
\vspace{-1.0cm}
\paragraph{}
Once you know this, every higher power of $L$ acting on $|v\rangle$ becomes:

\[
L^2 |v\rangle = L(L|v\rangle)
= L(\lambda |v\rangle)
= \lambda^2 |v\rangle,
\]

\[
L^3 |v\rangle = L(\lambda^2 |v\rangle)
= \lambda^3 |v\rangle,
\]
\vspace{-1.0cm}
\paragraph{}
and in general:

\[
L^k |v\rangle = \lambda^k |v\rangle.
\]
\vspace{-1.0cm}
\paragraph{}
so by replacing, we get the above transformation.
\vspace{-0.5cm}
\paragraph{}
\textbf{Result:} The eigenvector $|v\rangle$ remains unchanged, but the eigenvalue $\lambda$ is mapped to the phase factor $e^{i \lambda t}$.
\vspace{0.5cm}
\subsubsection{Trotter-Suzuki Decomposition}
 For non-commuting matrices $A$ and $B$, the standard exponential law fails:
\[
e^{A+B} \neq e^A e^B
\]

\vspace{-1.0cm}
\paragraph{}
This makes direct calculation of the matrix exponential computationally hard. To resolve this, we use the \textbf{Trotter-Suzuki decomposition} to approximate the exponential. By slicing the time $t$ into $n$ steps (Trotter steps), we can approximate the total operator as a product of exponentials of the individual terms:

\[
 e^{i(A+B)t} \approx \left( e^{iAt/n} e^{iBt/n} \right)^n
\]

\vspace{-0.5cm}
\paragraph{}
As $n \rightarrow \infty$, this approximation converges to the exact value, allowing us to implement the simulation using a sequence of simpler quantum gates.

\subsection{Quantum Phase Estimation (QPE).}

Quantum Phase Estimation aims to estimate the phase $\phi$ of an eigenvalue of a unitary operator $U$, defined by the eigenvalue equation:
\[
U|v\rangle = e^{2\pi i \phi} |v\rangle
\]
    
\vspace{-0.5cm}
\paragraph{}
where $\phi$ is a value between 0 and 1. We utilize two registers: an $m$-qubit "counting" register and the eigenvector  $|v\rangle$.

\paragraph{Step 1: Initialization}
We begin with the counting register in the all-zero state $|0\rangle^{\otimes m}$ and the target in $|v\rangle$:
\[
|\psi_0\rangle = |0\rangle^{\otimes m} \otimes |v\rangle
\]

\paragraph{Step 2: Superposition}
We apply the Hadamard gate ($H$) to all $m$ qubits in the counting register. This creates an equal superposition of all computational basis states:
\[
|\psi_1\rangle = \left( \frac{1}{\sqrt{2^m}} \sum_{k=0}^{2^m-1} |k\rangle \right) \otimes |v\rangle
\]
    
\vspace{-0.5cm}
\paragraph{}
Or, written in tensor product form for individual qubits:
\[
|\psi_1\rangle = \frac{1}{\sqrt{2}}\left(|0\rangle + |1\rangle\right) \otimes \dots \otimes \frac{1}{\sqrt{2}}\left(|0\rangle + |1\rangle\right) \otimes |v\rangle
\]

\paragraph{Step 3: Controlled-Unitary Operations (Phase Kickback)}
We apply a sequence of controlled-$U$ gates. The $j$-th qubit (from the bottom) controls the unitary $U^{2^j}$.
Due to the \textbf{Phase Kickback} phenomenon, applying a controlled-$U$ on eigenvector $|v\rangle$ leaves $|v\rangle$ unchanged but adds a phase $e^{2\pi i \phi}$ to the $|1\rangle$ component of the control qubit.
\vspace{-0.5cm}
\paragraph{}
For the first qubit applying $U^{2^0}$:
\[
\frac{1}{\sqrt{2}}(|0\rangle + |1\rangle) \xrightarrow{CU^{2^0}} \frac{1}{\sqrt{2}}(|0\rangle + e^{2\pi i \phi \cdot 2^0}|1\rangle)
\]

\vspace{-0.5cm}
\paragraph{}
For the $k$-th qubit applying $U^{2^k}$:
\[
\frac{1}{\sqrt{2}}(|0\rangle + |1\rangle) \xrightarrow{CU^{2^k}} \frac{1}{\sqrt{2}}(|0\rangle + e^{2\pi i \phi \cdot 2^k}|1\rangle)
\]

\vspace{-0.5cm}
\paragraph{}
After applying all controlled gates, the state of the counting register becomes:
\[
|\psi_{final}\rangle = \frac{1}{2^{m/2}} \left(|0\rangle + e^{2\pi i (2^{m-1}\phi)}|1\rangle\right) \otimes \dots \otimes \left(|0\rangle + e^{2\pi i (2^0\phi)}|1\rangle\right) \otimes |v\rangle
\]

\vspace{-0.5cm}
\paragraph{}
This can be rewritten mathematically as a summation:
\[
|\psi_{final}\rangle = \frac{1}{2^{m/2}} \sum_{k=0}^{2^m-1} e^{2\pi i \phi k} |k\rangle \otimes |v\rangle
\]

\paragraph{Step 4: Inverse Quantum Fourier Transform (IQFT)}

The QPE state before applying the inverse QFT is:

\[
\frac{1}{2^{m/2}} \sum_{k=0}^{2^m - 1} e^{2\pi i k \phi} \, |k\rangle .
\]
\vspace{-1.0cm}
\paragraph{}
This matches the definition of the Quantum Fourier Transform of an integer $x$:

\[
\mathrm{QFT}|x\rangle
= 
\frac{1}{\sqrt{2^m}}
\sum_{k=0}^{2^m - 1}
e^{2\pi i x k / 2^m} \, |k\rangle.
\]
\vspace{-1.0cm}
\paragraph{}
Compare the phase factors:

\[
e^{2\pi i k \phi}
\quad \text{vs} \quad
e^{2\pi i xk / 2^m}.
\]
\vspace{-1.0cm}
\paragraph{}
These expressions are equal when

\[
\phi = \frac{x}{2^m}
\quad \Longrightarrow \quad
x = 2^m \phi.
\]
\vspace{-1.0cm}
\paragraph{}
Thus, our state is exactly the Fourier transform of the integer $x = 2^m \phi$.
\vspace{-1.0cm}
\paragraph{}
Applying the inverse Quantum Fourier Transform gives:

\[
\mathrm{QFT}^\dagger \left(
\frac{1}{2^{m/2}}
\sum_{k=0}^{2^m - 1}
e^{2\pi i\phi k} \, |k\rangle
\right)
\approx |x\rangle
= |2^m \phi\rangle.
\]

\paragraph{Step 5: Measurement}
We measure the counting register to obtain the integer $x$. We then calculate the estimated phase:
\[
\phi \approx \frac{x}{2^m}
\]
\vspace{-1.0cm}
\paragraph{}

Finally, if we used a time parameter $t$ such that $\theta = \lambda t = 2\pi \phi$, we recover the original eigenvalue $\lambda$:
\[
\lambda = \frac{2\pi \phi}{t}
\]

\subsection{Why Do We Extract the Phase Values of the Graph Laplacian With Respect to a Superposition State?}

In the standard Quantum Phase Estimation (QPE) algorithm, a unitary operator $U$ acts on one of its eigenvectors $|v_k\rangle$, satisfying
\[
U |v_k\rangle = {\lambda_k} |v_k\rangle = e^{2\pi i \phi_k} |v_k\rangle,
\]
\vspace{-1.0cm}
\paragraph{}
where $\phi_k \in [0,1)$ denotes the eigenphase. In this case, the QPE circuit estimates the phase $\phi_k$ with high probability.

\vspace{-0.5cm}
\paragraph{}
However, in our hashing framework, the input to QPE is not a single eigenvector but a state defined in the node basis of the weighted graph. Each node of the graph corresponds to a computational basis state $\{|0\rangle, |1\rangle, \dots, |N-1\rangle\}$, and the overall input state can be represented as a superposition over these nodes:
\[
|\psi\rangle = \sum_{i=0}^{N-1} \alpha_i |i\rangle.
\]
Although we do not input the eigenvectors of the Laplacian directly, each node basis state $|i\rangle$ can itself be expressed as a linear combination of the Laplacian's eigenvectors $\{|v_k\rangle\}$:
\[
|i\rangle = \sum_k c_k^{(i)} |v_k\rangle.
\]
Thus, the input state $|\psi\rangle$ implicitly becomes a superposition of the Laplacian’s eigenvectors.
\[
|\psi\rangle = \sum_k c_k |v_k\rangle, \quad \text{where} \quad \sum_k |c_k|^2 = 1.
\]
Because QPE operations are linear, they act on each eigencomponent simultaneously. After applying the controlled-$U^{2^j}$ operations, the joint state of the phase and system registers becomes
\[
|\Psi\rangle = \frac{1}{2^{m/2}} \sum_{y=0}^{2^m-1} |y\rangle \otimes U^y |\psi\rangle
= \frac{1}{2^{m/2}} \sum_k c_k \sum_{y=0}^{2^m-1} e^{2\pi i \phi_k y} |y\rangle \otimes |v_k\rangle,
\]
Applying the inverse Quantum Fourier Transform (QFT) we get:
\[
|\Psi_{\text{after}}\rangle \approx \sum_k c_k \, |\tilde{\phi}_k\rangle_{\text{phase}} \otimes |v_k\rangle_{\text{system}}.
\]
This results in an entangled state between the estimated phase $\tilde{\phi}_k$ and its associated eigenvector $|v_k\rangle$.
When the phase register is measured, the state collapses to one of these branches. The probability of obtaining the phase value $\tilde{\phi}_k$ is proportional to $|c_k|^2$, and the system register simultaneously collapses to the corresponding eigenvector $|v_k\rangle$. Thus, QPE acts as a quantum spectral analyzer, decomposing the input state into its frequency (phase) components:
\[
P(\tilde{\phi}_k) \approx |c_k|^2.
\]
\vspace{-1.0cm}
\paragraph{}
If we run QPE many times, we see different phase values appear with different frequencies. These frequencies depend on the eigenvalues of 
U and how much the input state matches each eigenvector. This property is crucial in our hash construction.

\paragraph{How this mechanism prevents co-Spectral graphs from producing identical fingerprints?}
Although two graphs may share the same Laplacian eigenvalue multiset (i.e.,
they are \emph{co-spectral}), they generally possess different eigenvectors.
This is because the QPE sampling distribution depends on the overlaps
$|c_k|^2 = |\langle v_k | \psi \rangle|^2$ and not solely on the eigenvalues
$\{\lambda_k\}$, co-spectral graphs yield different phase distributions and
therefore distinct spectral fingerprints. Our hash is thus sensitive not only
to the spectrum but also to the eigenbasis structure, preventing collisions
between co-spectral but structurally different graphs.

\subsection*{Example: Representing node as a linear combination of the eigenvectors}
\subsubsection*{Three-Node Weighted Graph }

To illustrate how the Quantum Phase Estimation (QPE) algorithm extracts spectral fingerprints from a graph Laplacian, we consider a simple weighted graph $G$ with three nodes $\{v_1, v_2, v_3\}$, where the edge weights encode the connection strengths between nodes.
 
\begin{figure}[h!]
    \centering
    \includegraphics[width=1.0\textwidth]{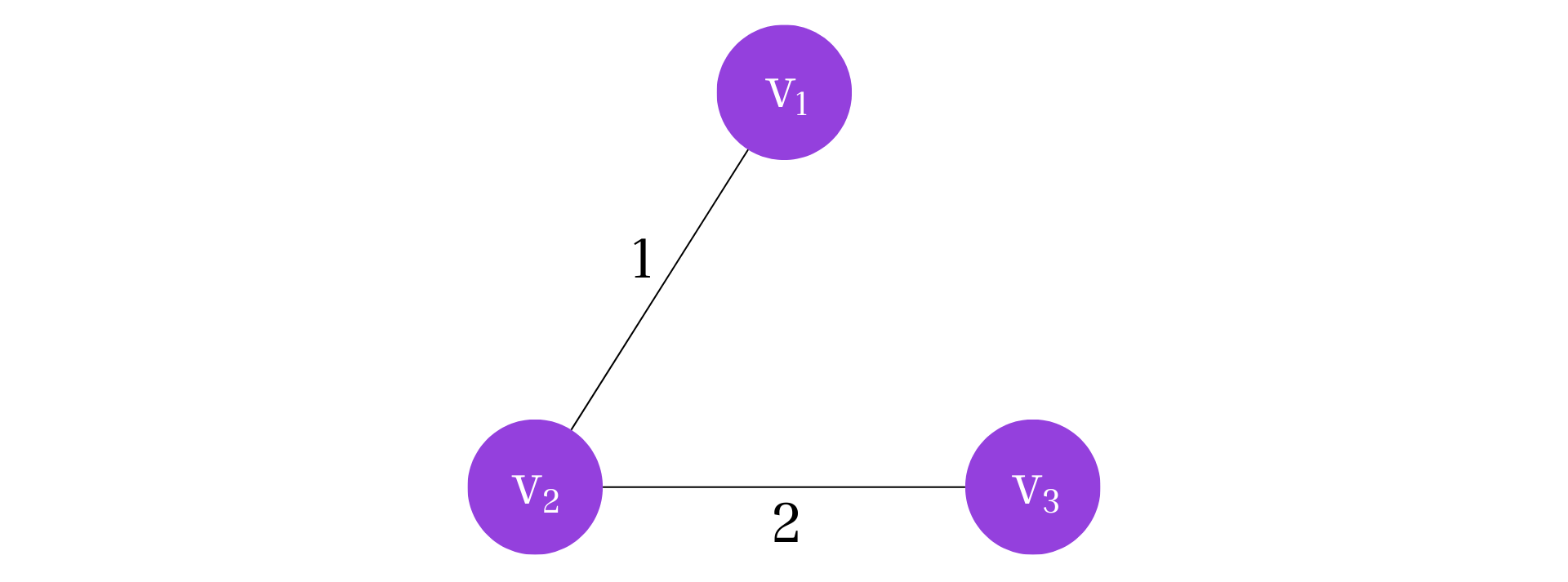}
    \caption{A simple 3-node weighted graph. Node $v_1$ is connected to $v_2$ with weight 1, and $v_2$ is connected to $v_3$ with weight 2. There is no direct edge between $v_1$ and $v_3$.}
    \label{fig:spectral_fingerprint}
\end{figure}
\vspace{-1.0cm}
\paragraph{}
The weighted adjacency and degree matrices of this graph are:
\[
A =
\begin{bmatrix}
0 & 1 & 0 \\
1 & 0 & 2 \\
0 & 2 & 0
\end{bmatrix},
\quad
D =
\begin{bmatrix}
1 & 0 & 0 \\
0 & 3 & 0 \\
0 & 0 & 2
\end{bmatrix}.
\]
Hence, the graph Laplacian is
\[
L = D - A =
\begin{bmatrix}
\ 1 & -1 & \ 0 \\
-1 & \ 3 & -2 \\
\ 0 & -2 & \ 2
\end{bmatrix}.
\]
\vspace{-1.0cm}
\paragraph{}
The Laplacian $L$ encodes all structural relationships in the graph: diagonal elements represent node degrees (total connection strengths), while off-diagonal terms represent negative edge weights. The eigenvalues and eigenvectors of $L$ define the graph’s intrinsic vibration modes (or diffusion modes).

\subsubsection*{Node Basis as a Superposition of Eigenvectors}
Let us see how the nodes can be represented as a linear combination of the eigen vectors
When we physically compute the eigenvalues and eigenvectors of $L$ yields approximately:
\[
\lambda_1 = 0.36, \quad
\lambda_2 = 2.20, \quad
\lambda_3 = 3.44,
\]
with corresponding normalized eigenvectors:
\[
|\phi_1\rangle =
\begin{bmatrix}
0.62 \\ 0.58 \\ 0.53
\end{bmatrix}, \quad
|\phi_2\rangle =
\begin{bmatrix}
\ 0.77 \\ -0.63 \\ -0.08
\end{bmatrix}, \quad
|\phi_3\rangle =
\begin{bmatrix}
\ 0.14 \\ \ 0.51 \\ -0.85
\end{bmatrix}.
\]
\vspace{-1.0cm}
\paragraph{}
Each node in the graph corresponds to a computational basis vector. For example, node $v_1$ corresponds to:
\[
|1\rangle = 
\begin{bmatrix}
1 \\ 0 \\ 0
\end{bmatrix}.
\]
This node-basis state can be expressed as a linear combination of the Laplacian’s eigenvectors:
\[
|1\rangle = c_1 |\phi_1\rangle + c_2 |\phi_2\rangle + c_3 |\phi_3\rangle,
\]
where the coefficients $c_k$ represent the overlap between $|1\rangle$ and each eigenvector:
\[
c_k = \langle \phi_k | 1 \rangle.
\]
Using the eigenvectors above, we find:
\[
c_1 = 0.62, \quad
c_2 = 0.77, \quad
c_3 = 0.14.
\]
Thus, the node $v_1$ is not associated with a single eigenmode but rather a weighted superposition of all three Laplacian eigenvectors. We see how when the node basis is used it doesn't relate to a single eigenvector but its superposition.

\subsection{Heat trace value}
Heat trace value is the exponentiated weighted average of the estimated eigenvalues,
\[
h(t)
= \sum_k |c_k|^2 e^{-t\tilde{\lambda}_k},
\]
where $\{\tilde{\lambda}_k\}$ is the estimated eigenvalue with respect to the superposition state (input to QPE), not the eigenvalues of the graph Laplacian.
\paragraph{Why the Exponential Appears in the Heat Kernel.}
The exponential function in the heat kernel is not a convention; it arises
directly from the physics of diffusion. The heat (or diffusion) equation on a
graph,
\[
\frac{\partial f}{\partial t} = -L f,
\]
is a first-order linear time-evolution equation whose generator is the graph
Laplacian $L$. Solving this differential equation yields
\[
f(t) = e^{-tL} f(0),
\]
so the operator $e^{-tL}$ describes how heat or probability diffuses across
the graph over time. When $L$ is diagonalized,
\[
L v_k = \lambda_k v_k,
\]
each eigenmode evolves independently as
\[
a_k(t) = e^{-\lambda_k t} a_k(0),
\]
showing that the eigenvalue $\lambda_k$ determines the exponential decay rate
of the corresponding mode. Thus, the heat trace
\[
Z(t) = \operatorname{Tr}(e^{-tL}) = \sum_{k} e^{-t\lambda_k}
\]
captures how all diffusion modes decay over time. The exponential appears
naturally because diffusion dynamics are governed by exponential decay of the
Laplacian's eigenmodes.

\subsection{Construction of the Spectral Fingerprint}
By evaluating the heat trace over multiple time scales $t_1, t_2, \ldots, t_T$, we construct the feature vector
\[
\mathbf{h} = [h(t_1), h(t_2), \ldots, h(t_T)],
\]
which we define as the spectral fingerprint of the graph. This fingerprint captures both eigenvalue-based and eigenvector-weighted characteristics, providing a unique, multiscale signature of the graph’s structure. It remains invariant under graph isomorphism while distinguishing non-isomorphic, cospectral graphs through their differing eigenvector contributions.

\section{Workflow - Message to Hash}\label{sec5}
The overall workflow of the proposed QGH-256 Algorithm is illustrated in Fig. 6. The process consists of several key stages that sequentially transform the input message into a high-entropy spectral fingerprint and finally into a fixed-length binary hash.
\vspace{-0.5cm}
\begin{figure}[h!]
    \centering
    \includegraphics[width=0.9\textwidth]{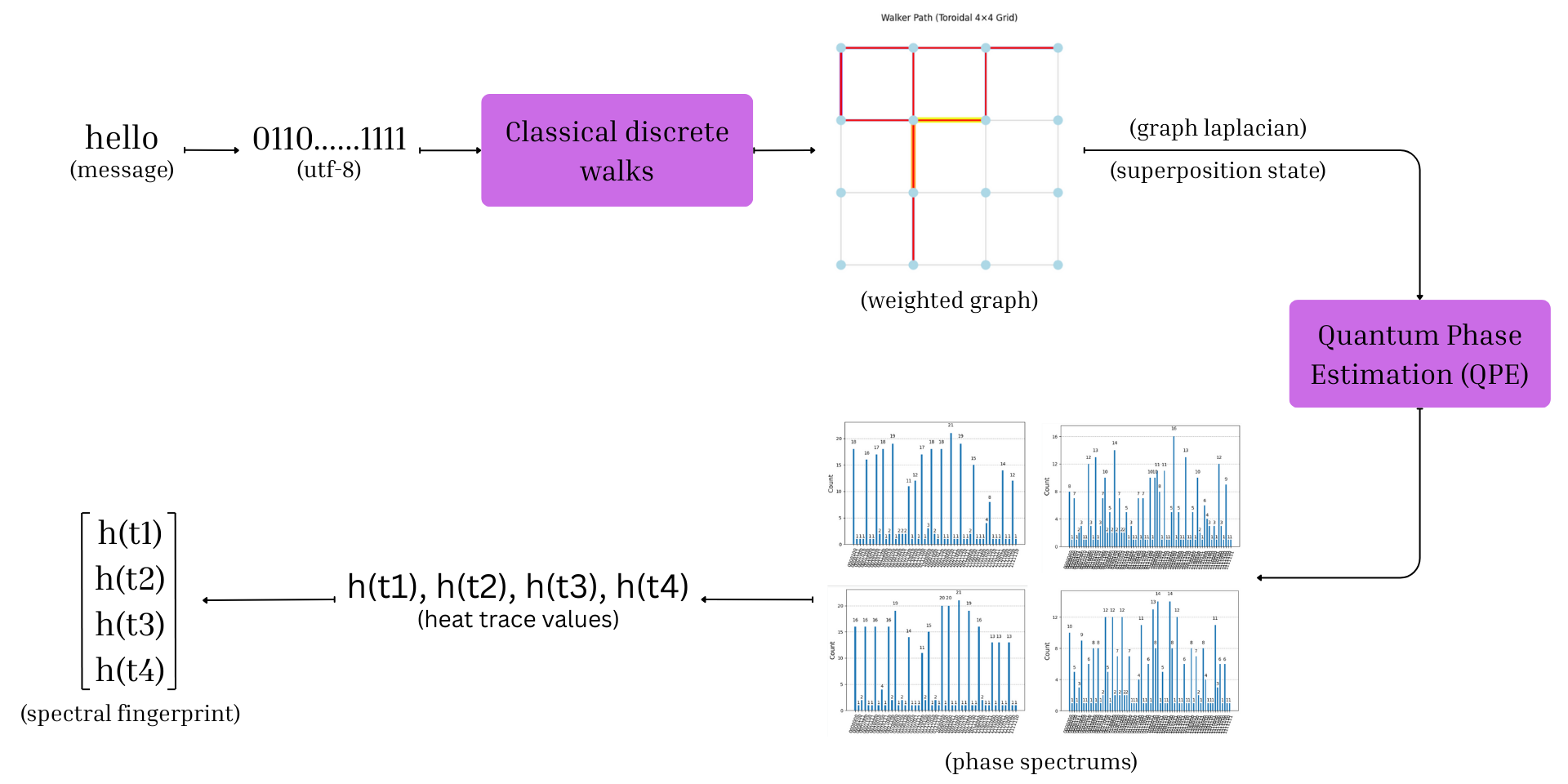}
    \caption{Overall workflow of the algorithm.}
    \label{fig:spectral_fingerprint}
\end{figure}
\FloatBarrier

\paragraph{}
\vspace{-0.5cm}

\begin{enumerate}
    \item \textbf{Message Encoding via Discrete Walk:}  
    The input message is first converted into UTF-8 and then mapped onto a weighted graph through classical discrete walker dynamics on a $4 \times 4$ toroidal grid. Each 2-bit symbol (00, 01, 10, 11) in the UTF-8 encoding specifies the walker’s direction, producing a message-dependent adjacency structure.

    \vspace{0.5cm}
    \item \textbf{Graph Laplacian Computation:}  
    From the constructed weighted graph, the Laplacian matrix $L = D - A$ is derived, where $D$ is the degree matrix and $A$ is the adjacency matrix. The Laplacian encodes both local connectivity and global structure of the graph.
    \vspace{0.5cm}
    \item \textbf{Quantum Phase Estimation (QPE):}  
    The Laplacian matrix is treated as the Hamiltonian operator for simulated quantum evolution. Using Quantum Phase Estimation, the eigenphase spectrum corresponding to the superposition input state is extracted. This spectrum captures essential graph spectral characteristics.
    \vspace{0.5cm}
    \item \textbf{Heat Trace Evaluation:}  
    For each time evolution step $t_i$, the exponentiated weighted average of the obtained phase values is computed to derive the \textit{heat trace} value $h(t_i)$. This process is repeated for multiple time steps to track spectral evolution over time.
    \vspace{0.5cm}
    \item \textbf{Spectral Fingerprint Generation:}  
    The sequence of heat trace values $\{h(t_1), h(t_2), \ldots, h(t_n)\}$ forms the spectral fingerprint of the graph. This fingerprint uniquely represents both the structural and dynamical properties of the message-encoded graph.
    \vspace{0.5cm}
    \item \textbf{Hash Construction:}  
    Finally, the spectral fingerprint vector is post-processed and converted into a fixed-length (256-bit) binary hash. This resulting hash serves as a compact, collision-resistant, and quantum-compatible digital signature of the input message.
\end{enumerate}

\section{Evaluation of the Cryptographic Properties of QGH-256}\label{sec6}
 The essential properties that define the strength and reliability of a cryptographic hash function include: 

\begin{enumerate}[label=\alph*)]
    \item Determinism
    \item Avalanche Effect
    \item Collision Resistance
    \item Pre-image Resistance
    \item Computation Time
\end{enumerate}

\vspace{-1.0cm}
\paragraph{}
In our proposed quantum hash function based on a message-induced classical random walk and spectral fingerprinting, we validated these core properties using the Seeded Qiskit StatevectorSimulator, which enables noiseless simulation of the underlying quantum circuit. The following sections discuss each property in detail along with the corresponding validation approach.

\subsection{A) Determinism}
\begin{figure}[h!]
    \centering
    \includegraphics[width=0.5\textwidth]{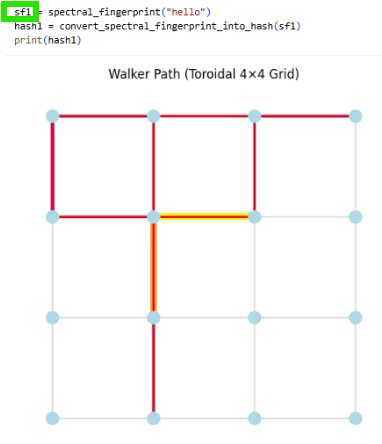}
    \label{fig:spectral_fingerprint}
\end{figure}

\begin{figure}[h!]
    \centering
    \includegraphics[width=0.5\textwidth]{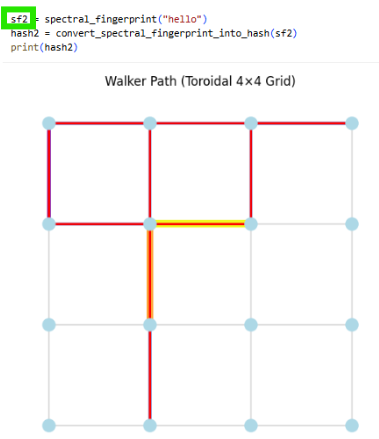}
       \caption{Code snippet showing same graphs both the times we hash "Hello", proving Deterministic behavior}

  \label{fig:spectral_fingerprint}
\end{figure}

\vspace{-0.5cm}
\paragraph{}
A hash function must be deterministic, meaning that for a given input message $M$, the output hash $H(M)$ should always be identical, regardless of how many times the hashing process is executed. Mathematically, this can be expressed as:

\[
H(M_1) = H(M_2) \quad \forall \, M_1 = M_2
\]
To validate determinism, the same message was hashed twice, and the \textit{Hamming distance} between the resulting bitstrings was calculated as:

\[
d_H(x, y) = \sum_{i=1}^{n} |x_i - y_i|
\]
where $x$ and $y$ are the two hash outputs, and $n$ is the bit-length of the hash. For a deterministic hash function, $d_H(x, y) = 0$.
\vspace{-1.0cm}
\paragraph{}
In our simulation, identical input messages consistently produced identical spectral fingerprints. The message-induced classical walker generates the same weighted graph structure for the same input, as the Laplacian and hence the eigenvalue spectrum remain  in an ideal quantum computer (simulation). Consequently, the spectral fingerprint, which serves as the quantum hash is unique and invariant for a given message, thereby confirming determinism.

\vspace{-0.5cm}
\paragraph{}
On real quantum hardware, however, noise can cause the measured hashes to differ across runs. With improved error-mitigation techniques, larger shot counts, and continued advances in quantum hardware, this variability can be reduced and eventually eliminatedcould be resolved.

\subsection{B) Avalanche Effect and C) Collision Resistance}\label{sec:collision}

\label{Collision}
A strong hash function must exhibit the \textit{avalanche effect}, where a minor change in the input message (even a single bit) leads to a significant and unpredictable change in the output hash. This ensures \textit{collision} doesn't occur.

\begin{figure}[H]
    \centering
    \includegraphics[width=0.5\textwidth]{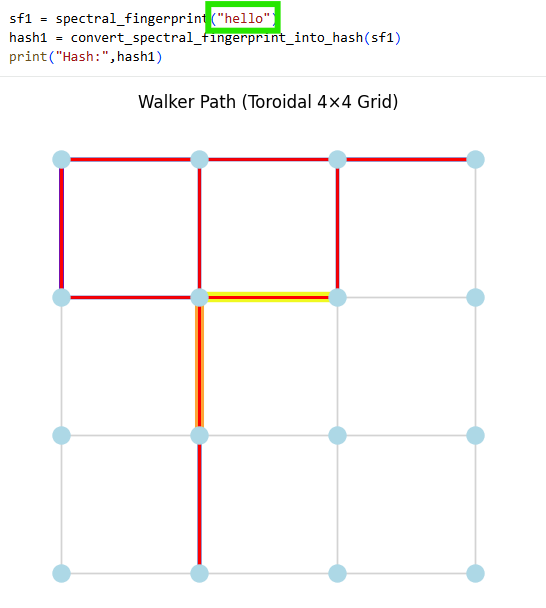}
    \caption{Graph produced by the message "hello"}
    \label{fig:hello_graph}

    \vspace{0.5cm} 

    \includegraphics[width=0.5\textwidth]{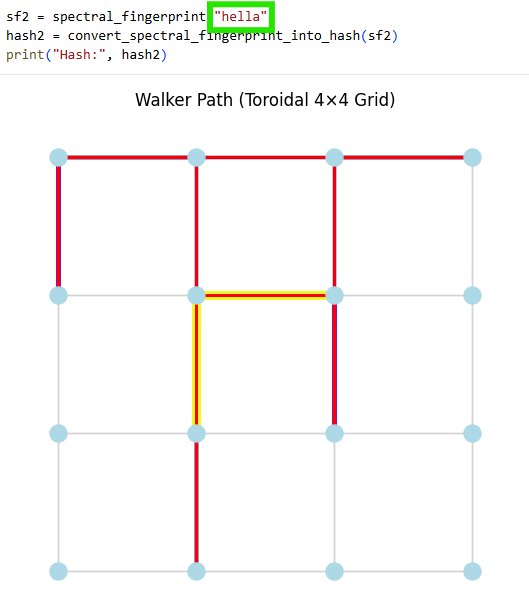}
    \caption{Graph produced by the message "hella"}
    \label{fig:hella_graph}
\end{figure}

\vspace{-1.0cm}
\paragraph{}
To verify this property, we generated hashes for the strings ``Hello'' and ``Hella'', differing by one character. The resulting hamming distance was significantly high, confirming the avalanche behavior. At the graph level, the bit difference alters the message-dependent random walk path, leading to a distinct graph and modified edge weights. This, in turn, changes the laplacian substantially, thereby amplifying small input differences into large output deviations.

\subsection{D) Pre-image Resistance}

Pre-image resistance ensures that given a hash value $H(M)$, it is computationally infeasible to reconstruct the original message $M$. In the context of our proposed method, the quantum hash is derived from the eigenvalue spectrum of the graph Laplacian, i.e.,
\[
H(M) = f(\text{Spec}(L(G_M)))
\]
where $\text{Spec}(L(G_M))$ denotes the set of eigenvalues (spectral fingerprint) of the Laplacian $L$ associated with the message-induced graph $G_M$.

\vspace{-0.5cm}
\paragraph{}
Reconstructing $G_M$ from its spectrum is known as the \textit{graph isospectrality problem}, which is computationally expensive to solve in general. Furthermore, even if an approximate reconstruction were possible, the graph being undirected introduces ambiguity as mentioned before in section \ref{msg_from_graph}, making it hard to infer the specific traversal order of bits corresponding to the input message. Thus, the mapping from message to graph, and subsequently from graph to spectral fingerprint, is effectively one-way and non-invertible. This ensures strong pre-image resistance.

\subsection{E) Computation Time}
In our algorithm, the QPE routine contributes a roughly constant computational
cost, while the classical walker stage scales with the length of the input
message. As the underlying graph is always a fixed $n \times n$ toroidal grid
($4 \times 4$ in our implementation), the size of the Laplacian matrix and
therefore the cost of the QPE stage remains constant and independent of the
message length. The only component whose runtime grows with the input size is
the classical discrete walker, since longer messages produce more 2-bit
movement steps and thus require more updates to the edge weights of the graph.

\begin{figure}[H]
    \centering
    \includegraphics[width=0.7\textwidth]{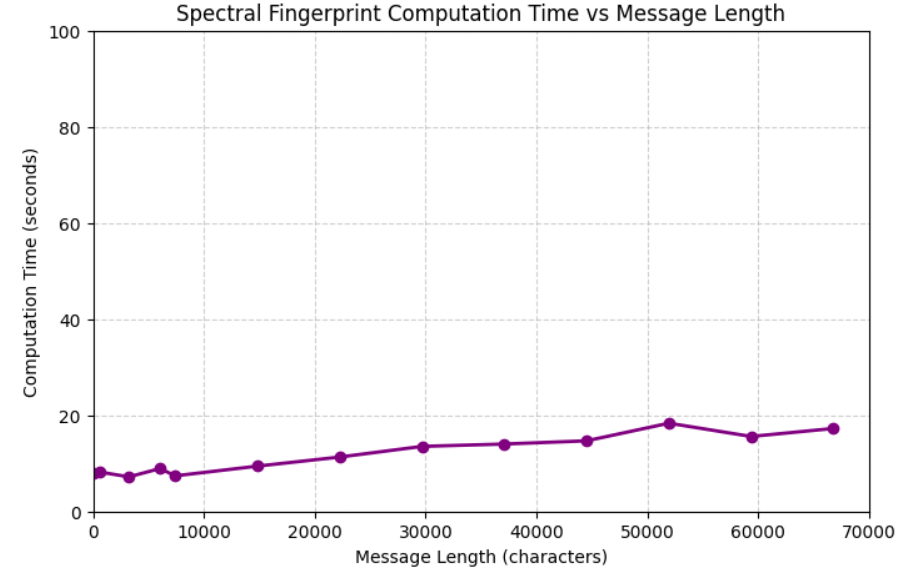}
    \caption{Graph between message length and computation time.}
\end{figure}

\section{Applications of the Proposed Hash Function}\label{sec7}

\subsection{Proof-of-Work Foundations}

\textbf{Hashcash}, proposed by Adam Back (1997), is a \textbf{proof-of-work (PoW)} system originally developed to combat spam and denial-of-service attacks. It requires senders to compute a hash satisfying a given difficulty condition (e.g., a fixed number of leading zeros). The computation is intentionally expensive, while verification remains trivial.

\vspace{-0.5cm}
\paragraph{}
In blockchain architectures, Hashcash principles were generalized to regulate block creation. Miners compete to discover a nonce that produces a block hash below a network-defined target, ensuring decentralized consensus and security.

\subsubsection*{Integration of the Spectral Hash}
The proposed \textbf{quantum-inspired spectral hashing algorithm} can replace conventional hash functions in PoW systems. Rather than relying on bitwise transformations, it constructs a \textbf{graph Laplacian} $L = A - D$ from input data and computes a \textbf{spectral fingerprint} via \textbf{quantum phase estimation (QPE)}.

\vspace{-0.5cm}
\paragraph{}
The aggregated sequence forms a high-entropy fingerprint that is collision-resistant and quantum-compatible. The mining difficulty is set by defining a target condition for the 256-bit spectral hash, similar to classical proof-of-work, but naturally resistant to quantum speed-ups.

\subsection{Data Integrity and Fingerprinting}

\textbf{Data integrity} ensures that information remains unaltered during storage or transmission. Conventionally, a cryptographic hash is computed on the data before transfer. The receiver recalculates the hash after download; any mismatch indicates corruption or tampering.
A practical example includes software-distribution websites that publish hash values (checksums) for downloadable applications. Users verify authenticity by recomputing the hash locally and comparing it with the official value. Another utilisation is in password protection
\vspace{-0.5cm}
\paragraph{}
\textbf{Fingerprinting} refers to generating a fixed-length digest that uniquely represents larger data. Because cryptographic hashes exhibit the \textbf{avalanche effect}, even a one-bit change drastically alters the fingerprint, enabling duplicate detection and authenticity verification.
\vspace{-0.5cm}
\paragraph{}
Our approach provides extreme sensitivity to input perturbations, high entropy. Thus, it functions as both a post-quantum integrity verifier and a high-dimensional data fingerprint.
\section{Quantum Threats to Classical Cryptography and Compliance with NIST PQC Standards}\label{sec8}

\subsection{NIST PQC Standards and Requirements}

Quantum computing represents a paradigm shift in computational science, leveraging 
the principles of quantum mechanics,such as superposition and entanglement,to solve 
problems that are computationally infeasible for classical systems \cite{nielsen2010, preskill2018}. 
This emerging technology introduces fundamentally new ways of processing information, 
enabling certain algorithms to outperform their classical counterparts by orders of 
magnitude. Among the most prominent examples are \textbf{Shor's algorithm} for factoring 
large integers and computing discrete logarithms \cite{shor1994}, and \textbf{Grover's algorithm} 
for unstructured search \cite{grover1996fast}. Both have direct implications for the foundational 
structures of modern cryptography.
\vspace{-0.5cm}
\paragraph{}
The exponential speedup offered by Shor’s algorithm threatens widely used 
\textbf{public-key encryption schemes} such as RSA etc
 \cite{bernstein2009}. In contrast, Grover’s quadratic speedup \textbf{reduces the effective 
security of symmetric-key algorithms}, including AES-128, by effectively halving their key strength 
\cite{moody2016}. These vulnerabilities could compromise the confidentiality, integrity, and 
authenticity of sensitive data across financial, governmental, military, and civilian digital 
infrastructures.
\vspace{-0.5cm}
\paragraph{}
The U.S. National Institute of Standards and Technology (NIST) has initiated the Post-Quantum Cryptography (PQC) Standardization process to identify algorithms resilient against quantum attacks. These standards are designed to replace or augment classical cryptographic primitives with quantum-resistant counterparts. 
\vspace{-0.5cm}
\paragraph{}
The NIST PQC framework focuses on four key criteria:

\begin{enumerate}
    \item \textbf{Security}
    \item \textbf{Performance}
    \item \textbf{Robustness}
    \item \textbf{Interoperability}
\end{enumerate}

\vspace{-0.7cm}
\paragraph{}
 NIST’s objective is to ensure a seamless transition to quantum-safe infrastructure across public and private sectors.

\subsection{Compliance of the QGH-256 with NIST PQC Standards}

The final step converts this spectral fingerprint into a 256-bit binary hash, representing a high-entropy, collision-resistant identifier of the original data. Unlike traditional hash functions that depend solely on linear bitwise operations, the spectral hash integrates both topological and dynamical information, substantially increasing resistance against both classical and quantum collision or pre-image attacks.
\vspace{-0.5cm}
\paragraph{}
This design achieves compliance with NIST PQC standards as follows:

\begin{itemize}
    \item \textbf{Quantum Resilience:} The spectral fingerprinting process involves complex functions of a graph Laplacian, which introduces non-linear dependencies between input and output. This significantly reduces vulnerability to Grover-type quadratic search attacks. Building a quantum oracle to compute this reversibly is very complicated and expensive.
    \vspace{0.5cm}
    \item \textbf{High Entropy and Collision Resistance:} As mentioned in \ref{sec:collision}, the properties of the graph construction and the Laplacian, ensure the algorithm is resistant to collisions.
    \vspace{0.5cm}
    \item \textbf{Interoperability:} The resulting 256-bit output maintains compatibility with existing blockchain and cryptographic infrastructures.
    \vspace{0.5cm}
    \item \textbf{Post-Quantum Efficiency:} Since the algorithm is quantum but classically executable through simulations, it provides forward security against near-term quantum threats while remaining efficient on current hardware.
\end{itemize}

\subsubsection{Grover's Algorithm and Its Implications in the proposed hash function}

Grover's algorithm is a foundational result in quantum computing that accelerates unstructured search. Given a black-box function $f:\{0,\dots,N-1\}\rightarrow\{0,1\}$ with a unique marked element, Grover's algorithm can identify the marked input using only $O(\sqrt{N})$ evaluations of $f$, offering a quadratic speedup over classical brute force search. This is often described metaphorically as ``finding a needle in a haystack'' using quantum interference: instead of checking each candidate sequentially, the algorithm amplifies the probability amplitude of the correct answer.
\vspace{-0.5cm}
\paragraph{}
A direct consequence is that security assumptions based on exhaustive search, such as preimage resistance of hash functions are weakened. For instance, a classical $2^{128}$-step brute-force search can in principle be reduced to approximately $2^{64}$ steps using Grover's algorithm. Hence, symmetric-key cryptosystems and hash functions are classified as ``quantum-vulnerable'', requiring increased output sizes or structural countermeasures.
\vspace{-1.0cm}
\paragraph{}
However, Grover's algorithm also incurs nontrivial implementation costs. It requires that the underlying function be represented as a reversible quantum circuit capable of acting on a superposition of all inputs. When the function being attacked is computationally complex, the cost per Grover iteration becomes extremely large, potentially offsetting the quadratic advantage.

\paragraph{Implications for the Proposed Quantum Hash Function.}
The algorithm proposed in this paper constitutes a \emph{heavy oracle} from the perspective of Grover's algorithm. The hash function proceeds as follows:
\begin{enumerate}
    \item Given a message $x$, construct a weighted graph on a fixed $4\times 4$ toroidal grid.
    \item Form the associated graph Laplacian $L = A - D$.
    \item Apply Quantum Phase Estimation (QPE) 
    \item Use these eigenvalues to compute heat traces, which constitute the final hash output.
\end{enumerate}
\vspace{-0.5cm}
\paragraph{}
Several components of this pipeline are intrinsically expensive on quantum hardware:
\begin{itemize}
    \item \textbf{QPE cost:} Quantum Phase Estimation requires multiple ancilla registers and deep controlled-unitary operations to achieve meaningful precision.
    \item \textbf{Matrix exponentiation:} Implementing $e^{-iLt}$ (required for QPE) involves Hamiltonian simulation of a graph Laplacian, which is resource-intensive even for modest matrix sizes.
\end{itemize}
\vspace{-0.5cm}
\paragraph{}
To apply Grover's algorithm for a preimage attack, an adversary must implement the entire hash function as a reversible quantum circuit. The resulting Grover oracle must:
\begin{enumerate}
    \item encode all possible messages $x$ in superposition,
    \item perform the full toroidal-grid mapping, Laplacian construction, and QPE-based spectral computation coherently on this superposition, and
    \item apply a phase flip conditioned on whether the computed spectral hash output equals a target hash value.
\end{enumerate}
Thus, while Grover's algorithm reduces the number of required trials from $N$ to $\sqrt{N}$, each trial becomes substantially more costly because it must simulate a deep quantum circuit implementing the entire spectral hashing pipeline. In effect, the algorithm's internal reliance on QPE and matrix exponentiation transforms it into a computationally heavy oracle, significantly increasing the per-iteration cost in a Grover-based preimage search. Consequently, although Grover's algorithm theoretically threatens any brute-force-based security assumption, the practical feasibility of such an attack against this quantum-inspired hashing method is constrained by the enormous circuit depth required.

\vspace{-0.5cm}
\paragraph{}
\vspace{-0.5cm}
\paragraph{}
In summary, QGH-256 satisfies the principal objectives outlined by NIST PQC: resistance to quantum attacks, strong entropy characteristics, and practical adaptability within secure communication and blockchain systems.


\section{Conclusion}\label{sec9}

In this work, we introduced a quantum hashing framework using discrete classical walker and Quantum Spectral Fingerprinting. The main outcomes
and directions of this study are summarized below:

\begin{itemize}
    \item \textbf{Core Contribution:}
We introduce a new quantum hash function that
combines a classical discrete walker with quantum spectral fingerprinting to
produce a compact, high-entropy hash. The message first influences the path of
a classical walker, which in turn shapes a weighted graph. We then apply
Quantum Phase Estimation (QPE) to the Laplacian of this graph. Unlike standard
QPE methods that operate on a single eigenvector, we input a superposition
state so that all eigenvectors contribute to the phase spectrum. This design
ensures that even co-spectral but non-isomorphic graphs produce different
spectral fingerprints, greatly reducing the chance of collisions and forming
the central novelty of our hashing approach.

    \vspace{0.5cm}
    \item \textbf{Evaluation on a seeded simulator:}
    On current Noisy Intermediate-Scale Quantum (NISQ) era quantum hardware, algorithms that rely on Quantum Phase
    Estimation (QPE) are highly sensitive to external noise. Even small amounts of
    decoherence or gate error introduce significant fluctuations in the extracted
    phase spectrum. These fluctuations affect the determinism required by our
    hashing pipeline: small variations in the measured phases propagate through the
    heat-kernel computation and alter the final fingerprint.
    \vspace{-0.5cm}
    \paragraph{}
    To suppress these fluctuations on real hardware, one would need to execute QPE with an extremely large number of shots in order to obtain a stable estimate of the phase distribution. However, increasing the shot count drastically inflates the total runtime, making the algorithm suitable for a future era of more advanced quantum hardware.
    \vspace{0.5cm}
    
    We evaluated our proposed algorithm using a seeded simulator, which produces
    deterministic and reproducible phase spectra by emulating an ideal, noiseless
    quantum evolution. This allows us to isolate and study the behavior of the
    construction without the interfering effects of hardware noise.

    \vspace{0.5cm}
    \item \textbf{Future Directions:}
    We aim to optimize the classical walker part to reduce preprocessing time and to implement the full algorithm on actual quantum hardware to evaluate its performance under realistic noise.

\end{itemize}

\vspace{0.5cm}
\clearpage
\backmatter

\nocite{*}
\bibliography{sn-bibliography}

\end{document}